\documentclass[10pt, conference, compsocconf]{IEEEtran}
\usepackage{amsmath}
\usepackage{algorithm}
\usepackage{algorithmic}
\usepackage{color}
\usepackage{stfloats}
\usepackage{supertabular,booktabs}
\usepackage{graphicx}
\usepackage{bm}
\usepackage{biblatex}
\begin{document}

\title{Stocks Vote with Their Feet: Can a Piece of Paper Document Fights the COVID-19 Pandemic?}
\author
{\IEEEauthorblockN{Jinhua Su \IEEEauthorrefmark{1} \IEEEauthorblockN{Qing Zhong \IEEEauthorrefmark{2}}}
\IEEEauthorblockA
{
\IEEEauthorrefmark{1}School of Statistics, Renmin University of China\\
\IEEEauthorrefmark{2}School of Finance, Renmin University of China\\
}

$ $\\
$\{2017201620,2017200183\}@ruc.edu.cn$
}

\maketitle

\begin{abstract}
Assessing the trend of the COVID-19 pandemic and policy effectiveness is essential for both policymakers and stock investors, but challenging because the crisis has unfolded with extreme speed and the previous index was not suitable for measuring policy effectiveness for COVID-19. This paper builds an index of policy effectiveness on fighting COVID-19 pandemic, whose building method is similar to the index of Policy Uncertainty, based on province-level paper documents released in China from Jan.1st to Apr.16th of 2020. This paper also studies the relationships among COVID-19 daily confirmed cases, stock market volatility, and document-based policy effectiveness in China. This paper uses the DCC-GARCH model to fit conditional covariance's change rule of multi-series. This paper finally tests four hypotheses, about the time-space difference of policy effectiveness and its overflow effect both on the COVID-19 pandemic and stock market. Through the inner interaction of this triad structure, we can bring forward more specific and scientific suggestions to maintain stability in the stock market at such exceptional times.
\end{abstract}


\IEEEpeerreviewmaketitle

\section{Introduction}
In early 2020, a pandemic called COVID-19 swept through China. In January-February, China's economy came to a standstill, and while maintaining social order and stable supplies, people stayed at home and made fighting the pandemic their first political task. Unfortunately, it become a global event\parencite{remuzzi2020covid}. 

At such exceptional times, the pandemic, the capital market, and government paper document, the triad interact to form unique institutional features.

The spread of the pandemic has prompted the government to take administrative and summoning measures to control the pandemic and to prevent its further deterioration. Likewise, once the government had published a piece of the paper document, the country deployed heavily from the top down to achieve the most vigorous form of resistance. At the same time, the pandemic had a huge impact on financial markets. Due to the fear of the pandemic, people worried about the synchronized slowdown of global economic growth\parencite{mckibbin2020global}, thus the risk of the Chinese New Year market suspension accumulated. Even pessimism and irrational behavior superimposed, finally, resulting in greater volatility in the Chinese stocks.\parencite{PetersonSpillover} 
Compared to the volatile impact of the pandemic on the capital market, the effect of the government policy was more explicit through social sentiment, while the effect to the capital market was more directly on the two driving forces, production and consumption, to stimulate the capital market.

Therefore, exploring the inner relationship among the pandemic, the stocks, and government documents can help to understand the economic and social phenomena in special times and can also study the effectiveness of Chinese government documents. To provide policy measures and development recommendations for reference when the global pandemic situation remains critical.

Indexes relevant to COVID-19 are most based on web-news data\parencite{dong2020interactive} while the policy document text contains valuable information but rarely being extracted. the Policy Uncertainty Index is measured by count specific terms relevant to uncertainty in the newspaper. As for COVID-19, there is no efficient index to measure corresponding policy effectiveness. Building Policy Effectiveness for COVID-19 is urgently needed. Once built, it could be used to analyze the triad relationship with the stock market and COVID-19 infection.

\section{Hypothesis}

Based on the principles of economics and political science, We put forward four hypotheses. The assumption of the effect of paper documents is based on China's modern administrative mechanism, government role, and government credibility, while the assumption of financial market fluctuations is based on the reality of China's financial market and the characteristics of the socialist market economy.
\subsection{Hypothesis 1: There is a lag of the corresponding policy promulgated between the local and the central.}

Policy Publishing in China is characterized by an "S-shaped" curve over time, in which the introduction of the same thematic policies by central and local governments goes through periods of slow, rapid, and steady proliferation. Spatially, it presents a variety of geographic diffusion effects such as proximity effect, hierarchical effect, axial effect, and agglomeration effect. However, considering the main body of action and the political  proliferation, the speed of response to the publication of central government documents varies somewhat across China.\parencite{Eva2014Policy}
\subsection{Hypothesis 2:There are regional differences in the management and response to the pandemic.}

China is a vast country, and factors such as geography, population, and transportation have led to different levels of pandemic severity throughout the country.\parencite{Shipan2012Policy} Wuhan, Hubei Province, is the earliest region in China where the pandemic was detected and the most strictly controlled. As for the southern and eastern parts of the country are the closest to Hubei compared to the far western and northeastern parts. The epidemic control is more stringent, comparing with other places.\parencite{Gilardi2010Who}
\subsection{Hypothesis 3:Fluctuations in the level of the pandemic affect the publication of government documents, while the power of the documents influences the next stage of the pandemic}

Whether the control of pandemic can be scientific, timely, and effective largely depends on the degree of specialization and decision-making capacity of the health sectors. \parencite{Gostin2002The}The relevance and forcefulness of the communication directly influence the next step of the pandemic.\parencite{Kostka2013Environmental}
\subsection{Hypothesis 4:The publication of the paper documents stimulates the stock market.}

The pandemic affected the country's economy on a large scale. It in turn sent shockwaves through stocks. In order to maintain financial stability and the normal operation of the capital market, the adoption of economic policies will stimulate liquidity in the financial market and reduce systemic risks.\parencite{Rosa2014The}
\section{Data}

As for this paper, we mainly use three types of time series data, two of which are easy to obtain, as listed in the following.

\begin{itemize}
\item \textbf{Stock Volatility}: We choose the SSE 50 index, compiled by the Shanghai Stock Exchange, daily price change to represent the stock market volatility in China.
\item \textbf{COVID-19 Severity}: The data we use is from the National Health Commission of China. Since suspected cases may highly correlate with confirmed cases, We only apply daily confirmed cases.
\item \textbf{Policy Effectiveness}: We crawled all paper document data from the province-level general office website in China. The data contains fields like release date, file number, title, and main text, where the earliest document is published in 1983. To meet our need of giving the focus on COVID-19, we finally use data in 2020(19252 observations). We use the LDA topic model to extract text data and build a dictionary about COVID-19 related policy, which will be demonstrated in the next section. Our paper document data and crawling code are available in GitHub\footnote{\href{https://github.com/JinhuaSu/Policy_crawler}{https://github.com/JinhuaSu/Policy\_crawler}}.

\end{itemize}

\section{Method}

In this section, We will first show you how to build the index of Policy Effectiveness on COVID-19, which can be promoted to a more general situation. Also, the series analysis method we use will be demonstrated a lot.

\subsection{Index building method}

Our goal here is to assess policy reflected to COVID-19 pandemic. A necessary first step is to quantify the effect and intensity of specific paper documents, which obviously give a response to current COVID-19 events, in a manner that delivers a suitable input into a statistical model. The Economic Policy Uncertainty Indices of
Baker, Bloom, and Davis\parencite{Baker2013Measuring} reflects the frequency of newspaper articles with one or more terms about “economics,” “policy” and “uncertainty” in roughly 2,000 U.S. newspapers. Their work inspired us to find keywords relevant to COVID-19 in paper document. We calculate the frequency of each word and build an LDA topic model to help us create a dictionary of those keywords.

\subsubsection{frequency analysis}
\begin{figure}[ht]\label{Figure 1}
\centering
\includegraphics[scale=0.1]{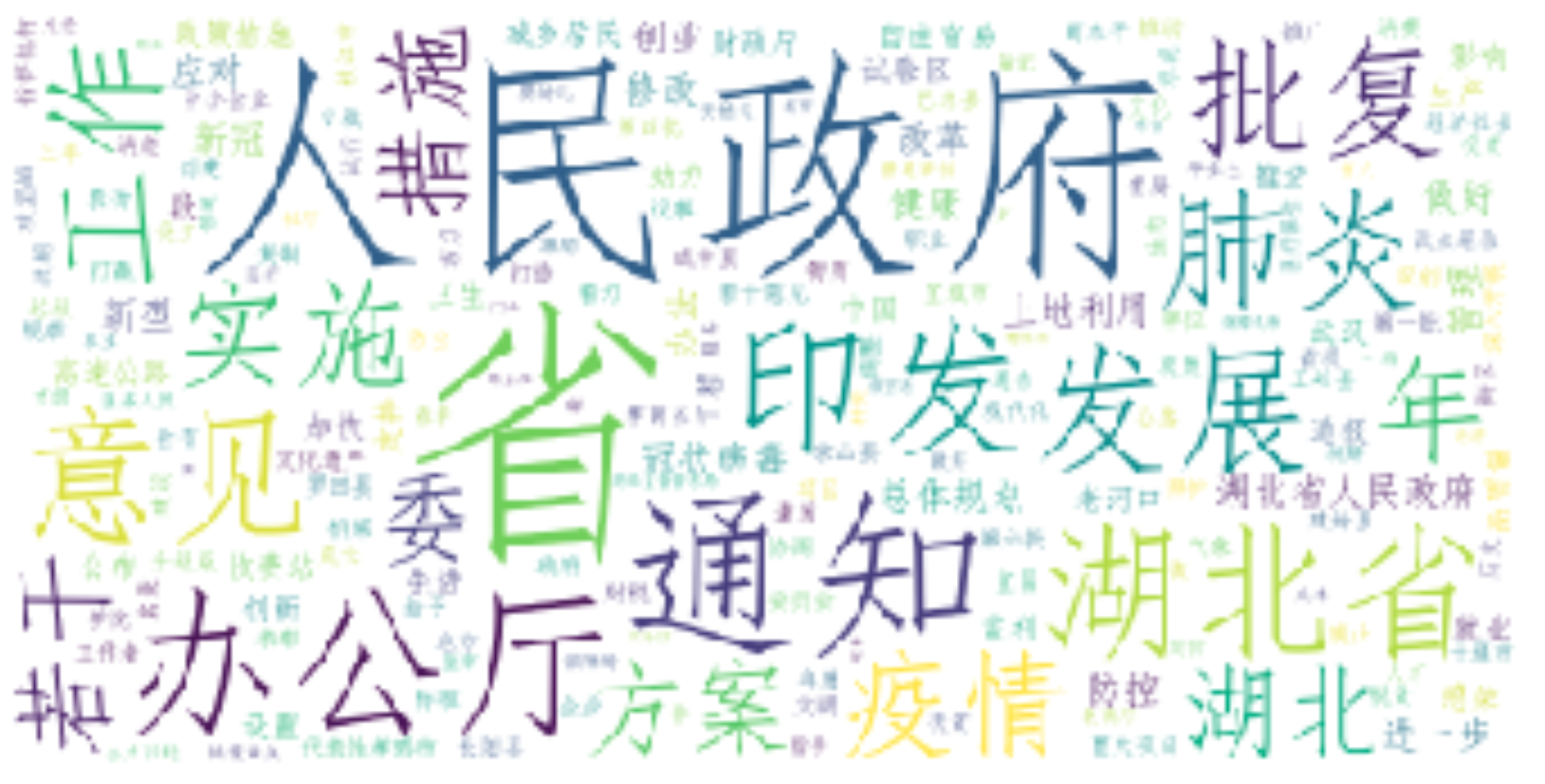}
\caption{Word cloud for paper documents in Hubei}
\end{figure}

\mbox{} \\
\indent To get a Chinese word list, since the Chinese don't have blank between words, we use Jieba package\footnote{\href{https://github.com/fxsjy/jieba}{https://github.com/fxsjy/jieba}} to cut each document. Afterword segmentation, the easiest way to describe the mathematical feature of text data is to count the words. We choose word cloud\ref{Figure 1} to visualize one of our results for Hubei province, where the outbreak happened. Frequent and relevant words will be chosen for our dictionary. Since different aspects of document will have different effect levels, variety and balance should be considered. That's why we built the LDA model to extract topics information in the following.

\subsubsection{LDA topic model}
\mbox{} \\
\indent In machine learning and natural language processing, a topic model\parencite{blei2003latent} is a type of statistical model for discovering the abstract "topics" that occur in a collection of documents. Topic modeling is a frequently used text-mining tool for the discovery of hidden semantic structures in a text body. 

We use the LDA model of Gensim to train my topic model. the latent Dirichlet allocation (LDA) is a generative statistical model that allows sets of observations to be explained by unobserved groups that explain why some parts of the data are similar. Finally we get five aspects for building our index, shown in the following table.

\begin{table}[h]
\caption{Key Words Type}\label{Table 1}
\centering
\begin{tabular}{c|c|p{4cm}}
\hline
{Sign} & {Word Type} & {Element}\\
\hline
A & Stringent Control & $0.040*child+0.032*care+0.015*home economics+0.009*nursery+0.010*pandemic$ \\
\hline
B & Transparent Information&$0.032*report+0.023*administration+0.015*complaint$\\
\hline
C & Construction Innovation & $0.012*pandemic+0.007*prevention+0.006*companies+0.005*contingency$ \\
\hline
D & Social Safety & $0.092*rescue+0.070*firefighting+0.065*emergency+0.030*team+0.016*forests+0.025*safe+0.019*production$ \\
\hline
E & economic support & $0.012*support+0.011*pandemic+0.011*loans+0.009*finance+0.009*guarantee$ \\
\hline
\end{tabular}
\end{table}

\subsubsection{Index building}
\mbox{} \\
\indent For each of the above five aspects, we choose high frequency and relevant word to build our dictionary of policy keywords for COVID-19 pandemic, which now contains 99 words. With the new-built dictionary, we can count these words appearing in each province website documents for each day.

Now we get our raw data, we will show how to calculate the Policy Effectiveness Index in this paper. Considering a series of effective policy, among relevant paper documents keywords will change with COVID-19 pandemic development. We suppose that some constant keywords appearing nearly in much more paper has less impact than the keywords only found in the document for a specific problem in the time of COVID-19.

Based on the above assumptions, we use the entropy weight method\parencite{qiyue2010structure} to weight the effectiveness of keywords in the dictionary built for COVID-19. With the following formula, we calculate the differentiation coefficient for each keyword. We finally build a series of daily policy effectiveness for each province.

\begin{equation}
p_{ij} = \frac{count_{ij}}{\sum_{i=1}^n count_{ij}}
\end{equation}

\begin{equation}
d_j = 1 + \frac{1}{ln(n)} \cdot  \sum_{i=1}^n p_{ij}ln(p_{ij})
\end{equation}

\begin{equation}
w_j = \frac{d_{ij}}{\sum_{i=1}^n d_{ij}}
\end{equation}

\begin{table}[h]
\caption{The differentiation coefficient of each keyword in new-built dictionary for COVID-19}\label{Table 2}
\centering
\begin{tabular}{p{1.6cm}|p{0.4cm}|p{1.7cm}|p{0.4cm}|p{1.5cm}|p{0.4cm}}
\hline
{word(type)} & {d} & {word(type)} & {d}&{word(type)} & {d}\\
\hline
nursery(A) & 0.69 &infant(A) & 0.70 &children(A) & 0.50 \\
transport(A) & 0.44 &vehicle(A) & 0.37 &access(A) & 0.39 \\
highway(A) & 0.41 &in-out(A) & 0.43 &shop(A) & 0.60 \\
supermarket(A) & 0.42 &up(A) & 0.56 &illegal(A) & 0.32 \\
inspection(A) & 0.34 &civil affairs(B) & 0.50 &mission(B) & 0.21 \\
rights(B) & 0.43 &administrative(B) & 0.53 &report(B) & 0.25 \\
underreport(B) & 0.43 &omission(B) & 0.44 &complaint(B) & 0.43\\
processing(B) & 0.28 &responsibility(B) & 0.22 &test(B) & 0.31 \\
performance(B) & 0.26 &science(C) & 0.23 &Stocks(C) & 0.39 \\
institutional(C) & 0.44 &company(C) & 0.40 &enterprise(C) & 0.17 \\
business(C) & 0.53 &management(C) & 0.33 &taxes(C) & 0.43 \\
structural(C) & 0.00 &greening(C) & 0.52 &farmers(C) & 0.42 \\
wood(C) & 0.45 &company(C) & 0.40 &ecology(C) & 0.36 \\
pollution(C) & 0.44 &cycle(C) & 0.46 &watersave(C) & 0.66\\
sewage(C) & 0.52 &agricultural(C) & 0.42 &mechanized(C) & 0.60 \\
farmland(C) & 0.60 &pesticides(C) & 0.50 &fertilizer(C) & 0.57 \\
employment(C) & 0.39 &employer(C) & 0.47 &worker(C) & 0.46 \\
vocational(C) & 0.51 &secondary(C) & 0.59 &education(C) & 0.48 \\
graduation(C) & 0.48 &recruitment(C) & 0.13 &meadow(D) & 0.72 \\
fire(D) & 0.48 &fight(D) & 0.61 &sandstorm(D) & 0.91 \\
weather(D) & 0.64 &disaster(D) & 0.57 &emergency(D) & 0.30 \\
check(D) & 0.48 &traffic(D) & 0.34 &highway(D) & 0.44 \\
party(D) & 0.44 &branch(D) & 0.49 &political(D) & 0.30 \\
grain-oil(D) & 0.54 &food(D) & 0.51 &electricity(E) & 0.42 \\
grid(E) & 0.52 &power(E) & 0.49 &toll road(E) & 0.54 \\
tolls(E) & 0.55 &rent(E) & 0.46 &rent free(E) & 0.21 \\
subsidies(E) & 0.38 &loan(E) & 0.37 &special(E) & 0.25 \\
credit(E) & 0.38 &financing(E) & 0.35 &liquidity(E) & 0.46 \\
cost(E) & 0.32 &platform(E) & 0.24 &security(E) & 0.43 \\
security(E) & 0.43&bank(E)&0.49&farming&0.46\\
sanctuary(E) & 0.56&food(E) & 0.47 &catering(E) & 0.41\\ store(E) & 0.48&return-work(E) & 0.39 &the-view(E)&0.53 \\
open(E) & 0.35&online(E) & 0.35 &posts(E) & 0.37\\
\hline
\end{tabular}
\end{table}

\subsection{Series Analysis method}

We use multiple series module to fit relevant time series in the field of policy, stock, and COVID-19. We use a common description method to show the series features, such as line chart, autocorrelation coefficient, and white noise test. Furthermore, we focus on the relation of the volatility of different series, we choose the DCC-GARCH(Dynamic Conditional Correlation - General Autoregressive Conditional Heteroskedasticity) model to fit the change rule of conditional covariance. The key formula of DCC is shown here, and we refer you to this paper of Engle\parencite{engle2001theoretical} for more details.

\begin{equation}
    Q_t = (1 - \alpha - \beta) \bar{Q} + \alpha u_{t-1}u_{t-1}^{'} + \beta Q_{t-1}
\end{equation}

For hypothesis 1, we use k-lag co-correlation between Central and each province to see whether there is a significant lag. As to hypothesis 2, we use 33 series of Policy Effective Index to fit DCC-GARCH and GO-GARCH model to show the variance series of six groups, which can be calculated by summing up all the elements of covariance matrix within a group. For hypothesis 3, we do a significance test on $\alpha$ and $\beta$ in the above fitted DCC-GARCH model. For hypothesis. For hypothesis 4, we fit a new DCC-GARCH model using Central Policy Index, COVID-19 confirmed cases, and SSE 50 Index, and then we plot the series of the fitted covariance matrix to analyze the relation of policy, COVID-19 and stock market.

\section{Result}

The property of the policy effectiveness index series will be shown first compared with the stock series and COVID-19 series. After testing their stationarity, We will use the DCC-GARCH model to fit them and draw some conclusions relevant to our hypothesis.

\subsection{Policy Effectiveness Index's property}

\begin{figure}[ht]\label{Figure 2}
\centering
\includegraphics[scale=0.2]{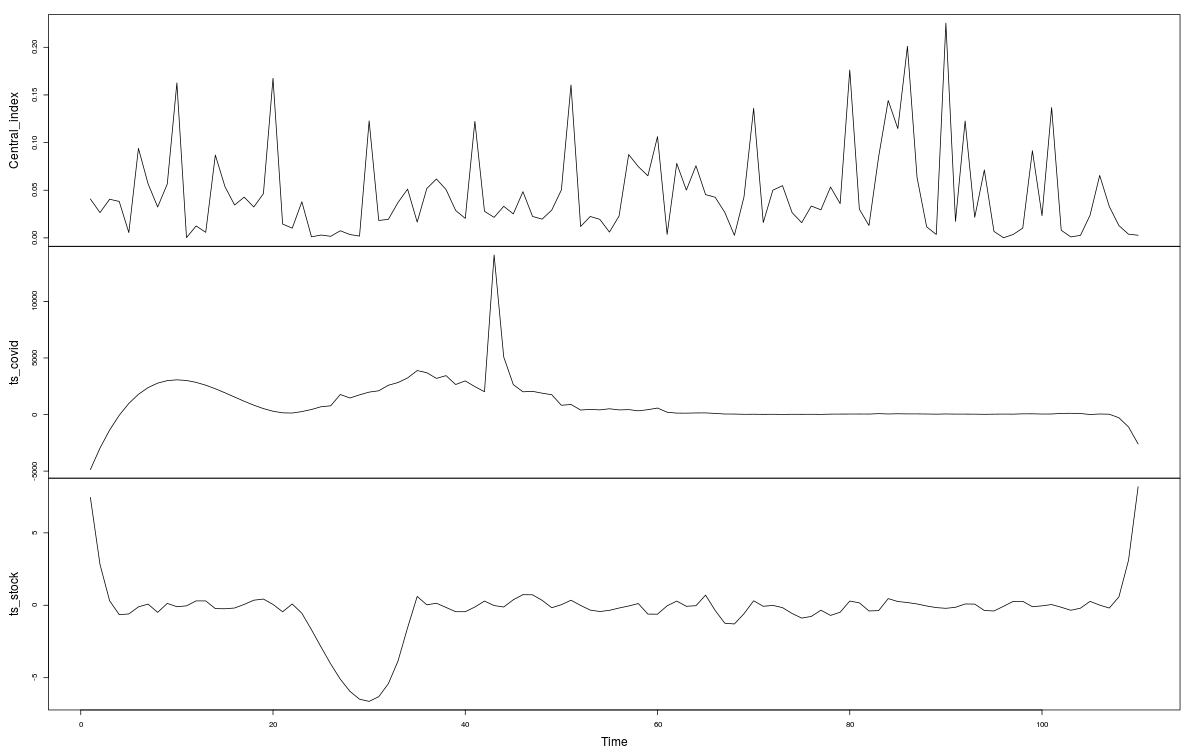}
\caption{Line charts of Central Policy Effectiveness Index, COVID-19 daily confirmed cases and SSE 50 Index daily change}
\end{figure}

In Figure 2\ref{Figure 2}, Policy Effectiveness Index appears to follow specific periodic change rule, while SSE 50 and COVID-19 series have an explosive jump in February. This time coincides with the Chines New year. During this period, the stock market closed, leading to virtually zero volatility. At the same time, the outbreak of pandemic in China reached its peak in February，especially, on February 13, the number of confirmed cases increases by more than 10,000 a day, which meets the cutting edge of the volatility curve. That is why we choose the heteroscedasticity model to fit them.

We apply the Augmented Dickey-Fuller Test to check their unit root. The results in table 3\ref{Table 3} all support stationarity hypothesis, which will avoid false regression in the model fitting part.

\begin{table}[h]
\caption{Unit Root Test(Augmented Dickey-Fuller Test)}\label{Table 3}
\centering
\begin{tabular}{p{2cm}|p{1.6cm}|p{1.6cm}|p{1.6cm}}
\hline
{Series} & {NCtype(p)} & {Ctype(p)} & {CTtype(p)}\\
\hline
Central Policy & -3.5538(0.01) & -6.8444(0.01) & -6.8275(0.01)\\
\hline
SSE 50 diff & -4.0176(0.01) & -4.0739(0.01) & -4.0501(0.01)\\
\hline
COVID-19 & -3,1984(0.01) & -3.8274(0.01) & -4.8969(0.01)\\
\hline
\end{tabular}
\end{table}

\subsection{Empirical analysis with DCC-GARCH}

In this part, we fit the DCC-GARCH model with multiple series of data and find some evidence supporting our four hypotheses.

\subsubsection{Test of Hypothesis 1}
\mbox{} \\
 \begin{equation}
     \rho(y,x,lag=k) = \frac{E[(y_t - Ey_t)(x_{t-k} - Ex_{t-k})]}{\sqrt{Var(y_t)Var(x_{t-k})}}
 \end{equation}

\begin{figure}[ht]\label{Figure 3}
\centering
\includegraphics[scale=0.2]{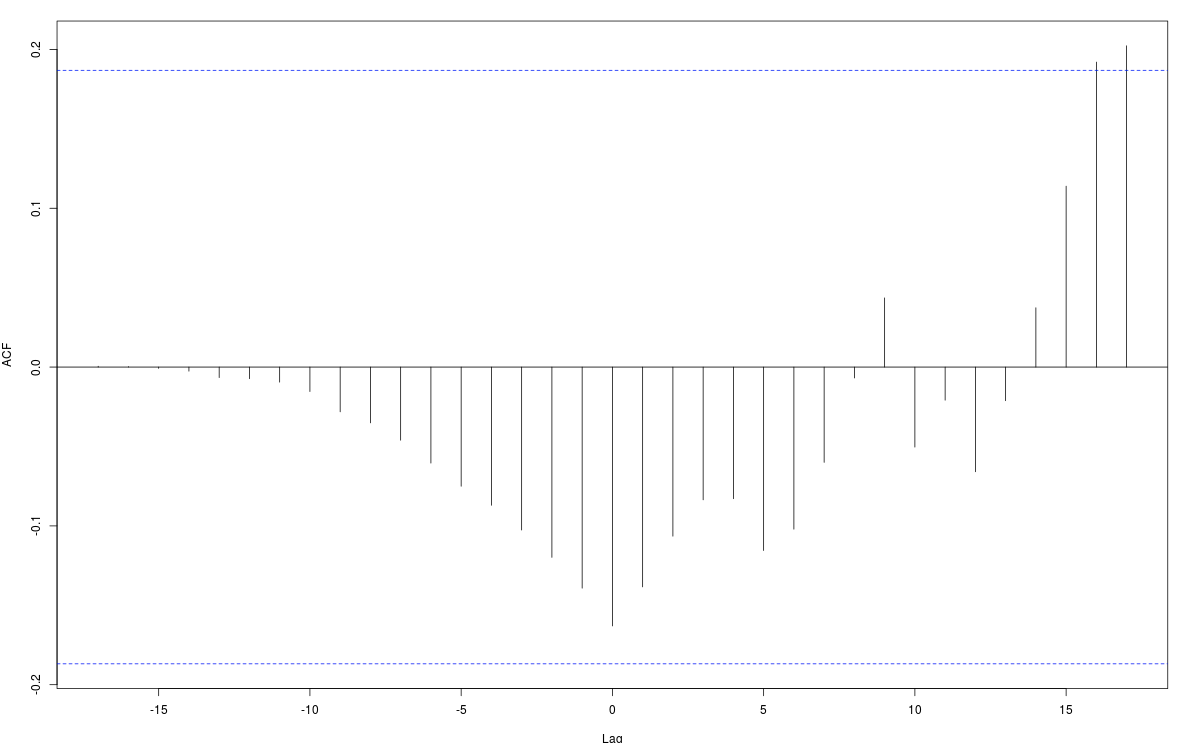}
\caption{k-lag Co-correlation coefficient between Hainan and Central}
\end{figure}

\indent Co-correlation coefficient is calculated by the above formula. Larger the value it is, the correlation between two series with a k-lag is more significant. Almost every policy information takes time to be reflected in the economy and society, and the central policy also takes time to be implemented by the local government. We draw 34 co-correlation graphs like \ref{Figure 3}{Figure 3} and find three features\ref{Table 4} appearing in most graphs. Documents published by the center need to be disseminated to local governments, which implement central advice and orders according to local conditions. This process takes some time. Due to the vast expanse of China and the wide variation in circumstances, there are differences in the time lag of paper documents. As seen in the Table 4\ref{Table 4}, when volatility lag is present, it is divided into strong and weak time lag. This meets the  "S-shaped" curve theory, and the special features bring a difference to the final time lags.

In the example of Hainan, we find explicit features like Right Volatility Bias, Short-time Negative Correlation, Long-time Positive Correlation. Hainan Province is located at the southern tip of China. When the pandemic first broke out in Hubei, the central government published an official document in response, and the Hainan provincial government remained in a wait-and-see state during the first 7 days. However, as the pandemic intensified, public policy was transmitted from central to local, so seven days later, the government published a similar policy in response to the central order. Thus, the curve of Hainan Province clearly shows the time lag of paper documents between the central and local.
 
\begin{table}[h]
\caption{Co-correlation property recognition and counting}\label{Table 4}
\centering
\begin{tabular}{p{1.5cm}|p{5cm}|c}
\hline
{Property} & {Description} & {Count}\\
\hline
Right Volatility Bias & When the k-lag co-correlation graph of a specific province has more significant volatility in the right part, there may a rule that the volatility of the specific province follows Central's with a lag. & 21\\
\hline
Short-time Negative Correlation & When lag $\leq$ 7, the co-correlation coefficient is negative. It may be the case that Central posed so many file while the specific province keep still at the beginning of pandemic. & 29\\
\hline
Long-time Positive Correlation  & If there exists a k-lag co-correlation coefficient with positive value double the s.e  When k $>$ 7, we can explain that it follow Central Policy with a k-day lag. & 18\\
\hline
\end{tabular}
\end{table}

\subsubsection{Test of Hypothesis 2}
\begin{figure}[ht]\label{Figure 4}
\centering
\includegraphics[scale=0.2]{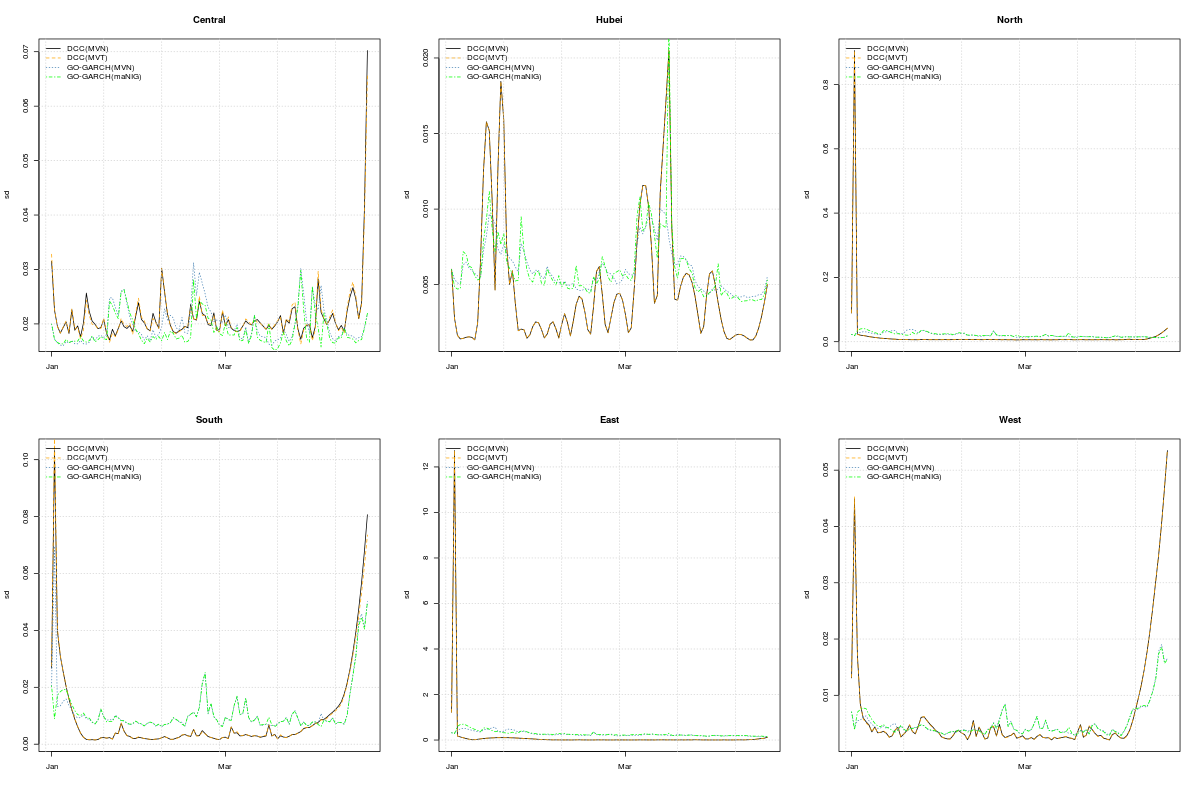}
\caption{Six group volatility series predicted by DCC-GARCH and GO-GARCH}
\end{figure}

\mbox{} \\
\indent We use 33 Policy Effectiveness Index series to fit different multiple Garch models, including DCC(E), DCC(T), GO-GARCH(MVN), GO-GARCH(maNIG). To show regional difference clearer, We define six groups, based on spatial position, and plot the volatility series\ref{Figure 4} by summing up corresponding elements in the predicted covariance matrix of  GARCH. Different regions in China seems to respond to COVID-19 in absolutely different manners. As can be seen in Figure 4\ref{Figure 4}, six figures represent the volatility of the Policy Effectiveness Index of central, Hubei, North, South, East, and west regions of China. The speed of publishing documents and the administrative power will affect the volatility. Based on this cause, Hubei as the first outbreak of the pandemic in China, the volatility fluctuated most violently. For example, in March, Hubei entered the most stringent control of the pandemic period, and the government took maximum efforts to combat the COVID-19. This moment relatively presented as the peak on the curve. From a geographical perspective, the south region is closest to Hubei, and with convenient transportation bringing large population movements, the South region has a faster and stronger rate of publication of paper documents than other regions.

\subsubsection{Test of Hypothesis 3}
\mbox{} \\
\indent Indeed reflexibility is a common phenomenon in macro-economy. In this part, we want to figure out how COVID-19, Policy Effectiveness, and Stock Emotion react to each other, from the perspective of volatility, by DCC-GARCH. First we apply a AR(1) on each series and then we fit a simple DCC-GARCH which only takes 1-lag information into consideration. The estimate and t-test of our model are shown in Table 5\ref{Table 5}. Each series contributes to volatility prediction and the joint part gives us a direct estimate of overflow effects. From the significant coefficients, it can be seen that the issuance of central documents is often not influenced by historical documents, but rather tends to take into account external factors such as the previous national situation and the development of the pandemic, while the pandemic has evolved as a cumulative result of historical circumstances in addition to other external factors. As for the stocks, it's condition and features are similar to the epidemic. The historical external factors of the triad are significant in the overflow effects of the triad interaction.

\begin{table}[h]
\caption{DCC-GARCH model estimate\cite{ding2001large}}\label{Table 5}
\centering
\begin{tabular}{ccccc}
\hline
{} &    {Estimate} & {Std. Error} &  {t value} & {$Pr(>|t|)$} \\
\hline
{$[Central].mu$}&{$0.046$}&{$0.004$}&{$11.06$}&{$0.000$}\\ 
{$[Central].ar1$}&{$-0.011$}&{$0.098$}&{$-0.12$}&{$0.904$}\\ 
{$[Central].omega$}&{$0.000$}&{$0.000$}&{$1.85$}&{$0.063$}\\ 
{$[Central].alpha1$}&{$0.039$}&{$0.052$}&{$0.74$}&{$0.457$}\\ 
{$[Central].beta1$}&{$0.829$}&{$0.063$}&{$13.06$}&{$0.000$}\\ 
{$[covid].mu$}&{$429.990$}&{$849.491$}&{$0.50$}&{$0.612$}\\ 
{$[covid].ar1$}&{$0.666$}&{$0.247$}&{$2.69$}&{$0.006$}\\ 
{$[covid].omega$}&{$0.000$}&{$0.000$}&{$0.00$}&{$1.000$}\\ 
{$[covid].alpha1$}&{$0.000$}&{$0.003$}&{$0.00$}&{$1.000$}\\ 
{$[covid].beta1$}&{$0.991$}&{$0.009$}&{$101.25$}&{$0.000$}\\ 
{$[stock].mu$}&{$-0.047$}&{$0.047$}&{$-1.00$}&{$0.315$}\\ 
{$[stock].ar1$}&{$0.266$}&{$0.050$}&{$5.31$}&{$0.000$}\\ 
{$[stock].omega$}&{$0.078$}&{$0.017$}&{$4.58$}&{$0.000$}\\ 
{$[stock].alpha1$}&{$1.000$}&{$0.291$}&{$3.42$}&{$0.000$}\\ 
{$[stock].beta1$}&{$0.000$}&{$0.033$}&{$0.00$}&{$1.000$}\\ 
{$[Joint]dcca1$}&{$0.313$}&{$0.135$}&{$2.31$}&{$0.020$}\\ 
{$[Joint]dccb1$}&{$0.290$}&{$0.250$}&{$1.16$}&{$0.245$}\\
\hline
\end{tabular}
\end{table}

\subsubsection{Test of Hypothesis 4}
\begin{figure}[ht]\label{Figure 5}
\centering
\includegraphics[scale=0.2]{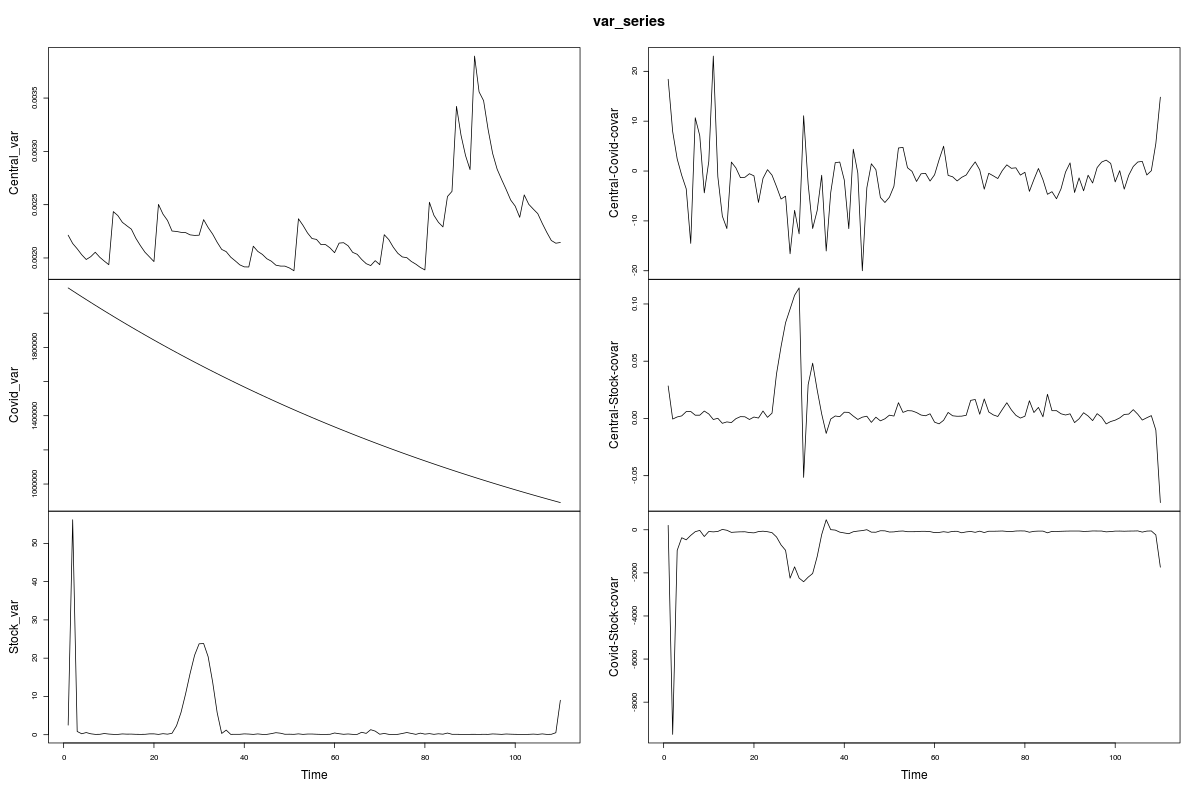}
\caption{Covariance matrix series of Central Policy Effectiveness Index, SSE 50 Index and COVID-19 daily confirmed cases}
\end{figure}

\mbox{} \\
\indent For a visual test of our hypothesis, We plot the variance and covariance of the above three series predicted by DCC-GARCH. As seen in Figure 5\ref{Figure 5}, in February, the average of central-COVID-covariance is negative, this suggests that, in February of the outbreak, the volatility of the policy effectiveness index is more stable, which means the control power with the pandemic is weak. Lack of timely and effective communication policies directly affect the next stage of the pandemic，as we can see in Figure 2\ref{Figure 2}, the COVID-29 achieved to the peak. Lessons learned, as the paperwork followed the development situation of epidemic, volatility covariance reverted to nearly zero. This explains the interaction between the epidemic and official effectiveness visually. The overflow effects of paper documents and the stock market are the opposite. Also in February, the central-stock-covariance is positive. This means that the stronger the paper document, the more volatile the stock market, that is, the paper document stimulates the stock market. This may be due to the government's publication of more economic policies related to releasing liquidity in the financial markets and supporting returning to work.

\section{Conclusion}
A system of policy effectiveness index made by keywords was constructed to extract data on the national public documents from January to April. It quantifies the effectiveness of the documents by the volatility of the indicators. After analysis, we empirically concluded that the policy effectiveness of central government documents is highly significant, although the time lag and power between central and local varies. The government has published six major policies on "stringent Control", "Transparent information", "construction innovation", "Social Safety" and "Economic Support" to control the pandemic and stabilize the economy. Just a piece of paper document can give the stock market a shot in the arm from the social emotions and economic base, while attracting the source of living water to the capital market.


\printbibliography

\end{document}